\documentclass[twocolumn,aps,showpacs,floats,letterpaper,floatfix, eqsecnum, groupedaddress]{revtex4-1}
\usepackage{graphicx}
\usepackage{dcolumn}
\usepackage{amsmath}
\usepackage{amsfonts}
\usepackage{amssymb}
\usepackage{epsfig,float,afterpage,wrapfig,psfrag}
\usepackage{natbib}

\newcommand{\beq}{\begin{equation}}
\newcommand{\eeq}{\end{equation}}
\newcommand{\bes}{\begin{subequations}}
\newcommand{\ees}{\end{subequations}}
\newcommand{\bea}{\begin{eqnarray}}
\newcommand{\eea}{\end{eqnarray}}
\newcommand{\ba}{\begin{array}}
\newcommand{\ea}{\end{array}}
\newcommand{\beqn}{\begin{eqnarray*}}
\newcommand{\eeqn}{\end{eqnarray*}}

\newcommand{\f}[2]{\frac{#1}{#2}}
\newcommand{\g}{\gamma}

\newcommand{\G}{\Gamma}
\newcommand{\la}{\langle}
\newcommand{\dg}{\dagger}
\newcommand{\ra}{\rangle}

\def\b0{{\bf 0}}

\def\nn{\nonumber}

\begin{document}
\title{Statistics of scattered photons from a driven three-level emitter in a one-dimensional open space}
\author{Dibyendu Roy$^1$ and Nilanjan Bondyopadhaya$^2$} 
\affiliation{$^1$Theoretical Division and Center for Nonlinear Studies, Los Alamos National Laboratory, Los Alamos, New Mexico 87545, USA} 
\affiliation{$^2$Integrated Science Education and Research Centre, Visva-Bharati University, Santiniketan, WB 731235, India}
\begin{abstract}
We derive the statistics of scattered photons from a $\Lambda$- or ladder-type three-level emitter (3LE) embedded in a 1D open waveguide. The weak probe photons in the waveguide are coupled to one of the two allowed transitions of the 3LE, and the other transition is driven by a control beam. This system shows electromagnetically induced transparency (EIT) which is accompanied with the Autler-Townes splitting (ATS) at a strong driving by the control beam, and some of these effects have been observed recently. We show that the nature of second-order coherence of the transmitted probe photons near two-photon resonance changes from bunching to antibunching to constant as strength of the control beam is ramped up from zero to a higher value where the ATS appears.
\end{abstract}

\pacs{03.65.Nk, 42.50.Ar, 42.50.Gy, 85.25.Cp}
\vspace{0.0cm}
\maketitle
\section{Introduction}
\label{intro}
Strong light-matter interaction in open space at the level of single atom and a few photons can be created by coupling a real or artificial atom to photon modes confined in an open one-dimensional (1D) waveguide \cite{Shen07, Chang07, Yudson08, Zumofen08, Shi09, Roy10, Liao10, Zheng10, Witthaut10, Roy11, Gerhardt07, Wrigge08, Tey08, Hwang09, Astafiev10, Astafiev10b, Abdumalikov10, Hetet11, Hoi11, Hoi12, Zheng12, Hoi13}. Efficient strong coupling between matter and photon field has been achieved by using highly confined propagating microwave photon modes in a 1D open superconducting transmission line and a large dipole moment of an artificial atom such as a superconducting qubit \cite{Astafiev10, Astafiev10b, Abdumalikov10, Hoi11, Hoi12, Hoi13}. A destructive interference between the emitted photons from a two-level atom and the incident photons in the waveguide yields extinction of the transmitted photons for the atom being side-coupled to a weak incident photon field. Extinction efficiencies greater than $99\%$ have been observed in recent experiments with superconducting transmission lines and superconducting `transmon' qubits \cite{Hoi11, Hoi12, Hoi13}. Other systems which are currently under extensive studies include surface plasmons of a metallic nanowire coupled to quantum dots or nanocrystals \cite{Akimov07} and line-defects in photonic crystals coupled to quantum dots \cite{Faraon07, Fushman08}.

\begin{figure}
\includegraphics[width=7.0cm]{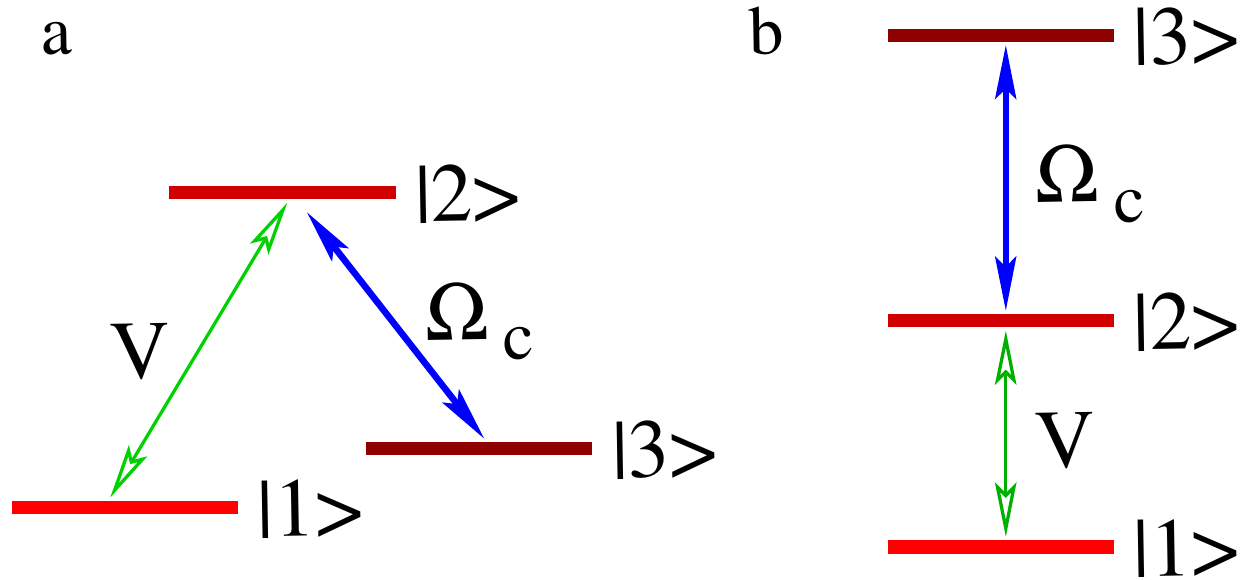}
\caption{Schematic of (a) $\Lambda$- and (b) ladder-type three-level emitters whose one of the two allowed transitions is coupled to probe photons by strength $V$ and the other transition is driven by a control beam with Rabi frequency $\Omega_c$.}
\label{sch}
\end{figure}

One of us (D.R.) has recently studied single- and two-photon scattering of a weak probe beam by a driven $\Lambda$-type three-level atom or emitter (3LE) which is embedded in a 1D open waveguide \cite{Roy11}. The excited state $|2\ra$ of the emitter (see Fig.\ref{sch}(a)) is connected to the state $|3\ra$ by a classical laser beam with Rabi frequency $\Omega_c$. We set energy of the ground state $|1\ra$ to be zero. Thus we can write the Hamiltonian of the 3LE within the rotating-wave approximation as $\mathcal{H}_{3LE}=(E_2-i\gamma_2/2)|2\ra\la2|+(E_2-\Delta-i\gamma_3/2)|3\ra\la3|+(\Omega_c/2)(|3\ra\la2|+|2\ra\la3|)$, where spontaneous emission loss from the 1D waveguide is modeled by including an imaginary part $-i\gamma_2/2$ and $-i\gamma_3/2$ to the energy of the respective states $|2\ra$  and $|3\ra$. The states $|1\ra$ and $|3\ra$ can be two hyperfine split states, and the transitions $|1\ra-|2\ra$ and $|2\ra - |3\ra$ would couple to different polarizations of light by selection rule. A probe beam in %the photon modes of 
the waveguide is sent near resonant to the transition $|1\ra-|2\ra$. We also consider that there is no direct transition between the states $|1\ra$ and $|3\ra$ by selection rule. An exact single-photon scattering state of the probe beam and the corresponding transmission line-shape showing electromagnetically induced transparency (EIT) \cite{Fleischhauer05} were calculated for this system in Ref.\cite{Witthaut10}. One of us has derived two-photon scattering state of the probe beam in this system for a weak control field and studied scaling of EIT line-shape for single and two probe photons \cite{Roy11}. Incident photons are in Fock-state in both the previous studies. 

In a set of recent experiments \cite{Abdumalikov10, Hoi11} it has been claimed to observe EIT line-shape for a driven ladder-type three-level superconducting qubit embedded in an open transmission line. It is also shown in these experiments that the Autler-Townes splitting (ATS) \cite{Autler55} appears when the one of the two allowed transitions of the qubit is driven by a strong control beam \cite{Abdumalikov10, Hoi11}. One interesting feature of a system of an emitter coupled to an open waveguide is strong photon-photon interactions generated by the two- or multi-level emitter by preventing multiple occupancy of photons locally at the emitter. It is known that scattered states from a two-level emitter embedded in a 1D waveguide can be non-classical \cite{Shen07, Roy10, Zheng10}. The antibunching of the reflected photons and superbunching of the transmitted photons have been recently demonstrated by measuring second-order coherence of the scattered fields from a two-level emitter \cite{Hoi12, Peropadre13}. However the nature of photon-photon correlations of the scattered probe beam from a 3LE in the presence of an arbitrary strong control beam has not been investigated in experiments \cite{Zheng12}. 

Photon-photon correlations induced tunneling or blocking of photons would affect  transmission and reflection of photons in the 3LE-waveguide systems. Therefore, a clear understanding of the nature of photon-photon correlations is very important to figure out functionality of various rudimentary quantum devices, such as a switchable mirror or a single-photon router using these systems \cite{Abdumalikov10, Hoi11}. A single-photon router can route a single-photon signal from an input port to either of two output ports. These devices might have important applications in building photonic quantum networks for quantum information processing. The emitted photons from the emitters in free-space can maintain the same envelope as the incident coherent state with a wide bandwidth limited only by the emitters' relaxation rate \cite{Hoi12}. This is an advantage of free-space strong light-matter interactions over the cavity mediated one since the properties of emitted photons from a cavity are limited by the cavity width and stochastic release by the cavity. Here we study scattering of multiple probe photons from a $\Lambda$-type 3LE for a general value of $\Omega_c$. In particular, we derive exact two- and multi-photon scattering states of the probe beam for an arbitrary $\Omega_c$ and calculate second-order coherence of the reflected and transmitted probe photons. The article is organized as follows. In Sec.\ref{EITATS} we discuss EIT and the ATS in single-photon transmission line-shape through a $\Lambda$- and a ladder-type 3LE. The nature of second-order coherence of the transmitted and reflected photons is studied in Sec.\ref{2OC}. We conclude with a short discussion in Sec.\ref{conl}. We provide a summary of all the technical details that are relevant to the main results in Appendices \ref{app1} and \ref{app2}. The Appendix \ref{app1} also provides explicit expression of the multi-photon scattering states of the probe beam.

%However we are not aware of any theoretical or experimental study on photon-photon correlations of the scattered probe beam from a 3LE in the presence of an arbitrary strong control beam. Therefore, we here study scattering of multiple probe photons from a $\Lambda$-type 3LE for a general value of $\Omega_c$. In particular, we derive exact two- and multi-photon scattering states of the probe beam for an arbitrary $\Omega_c$ and calculate second-order coherence of the reflected and transmitted probe photons.

\section{Electromagnetically induced transparency}
\label{EITATS}
The scattering of photons from a driven 3LE embedded in a 1D photonic waveguide can be described by the following Hamiltonian \cite{Roy11}
\bea
\mathcal{H}=\mathcal{H}_{wg}+\mathcal{H}_{3LE}+\mathcal{H}_c,\label{Ham}
\eea
where $\mathcal{H}_{wg}$ represents free probe photons in the waveguide and $\mathcal{H}_{3LE}$ for a driven 3LE is already introduced. The local coupling of the probe photons with the 3LE is given by $\mathcal{H}_c$. We consider a linear energy-momentum dispersion ($E_k=v_gk$) for the free probe photons, and divide the positive and negative momentum photons as right- and left-moving modes. Thus we write 
\bea
 \mathcal{H}_{wg}=-iv_g \int dx [a^{\dg}_R(x)\partial_x a_R(x)- a^{\dg}_L(x)\partial_x a_L(x)],
\eea
where $v_g$ is the group velocity of the photons and $a_R(x)~ [a_L(x)]$ is the annihilation operator of a right-(left-) moving photon at position $x$. 
In our model the 3LE is side-coupled to the propagating light fields locally at $x=0$; thus we write
\bea
\mathcal{H}_c=V|2\ra\la1|(a_R(0)+a_L(0))+h.c.,
\eea
where $V$ is coupling strength between the emitter and the probe photons. We set here $v_g=\hbar=1$.
 
\begin{figure}[htb]
\includegraphics[width=8.5cm]{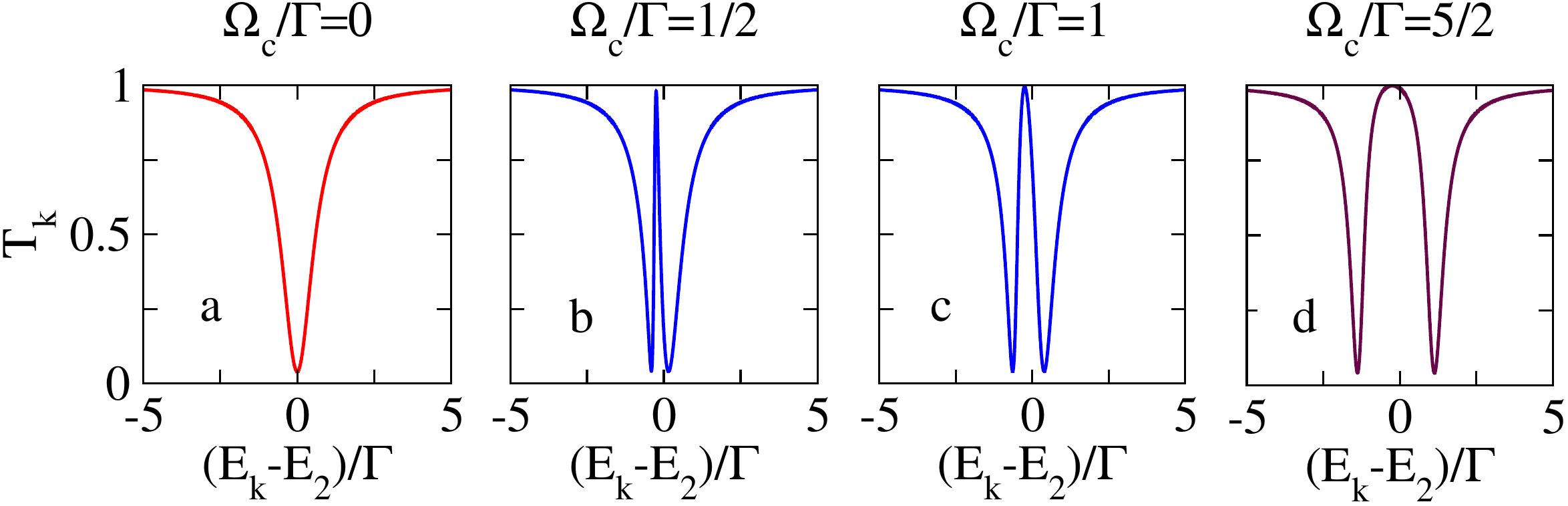}
\caption{Appearance of electromagnetically induced transparency (EIT) at two-photon resonance, $E_k-E_2=-\Delta$ when a weak control beam ($\Omega_c<\Gamma$) is switched on, and the Autler-Townes splitting (ATS) appears at a relatively strong control beam $(\Omega_c>\Gamma)$. The splitting between the Autler-Townes doublet is $\Omega_c$. The parameters are $\Delta/\Gamma=\g_2/\Gamma=1/4,~\g_3/\Gamma=1/40$ for a $\Lambda$-type emitter.}
\label{EIT1}
\end{figure}

The single-photon transmission and reflection line-shapes for a driven $\Lambda$-type 3LE coupled to a 1D waveguide have been reported earlier in Refs.\cite{Witthaut10, Roy11}. The single-photon transmission and reflection amplitudes are given respectively by $\tilde{t}_{k}=(t_k+1)/2=\chi/(\chi+i\Gamma/2)$ and $\tilde{r}_{k}=(t_k-1)/2=-0.5i\Gamma/(\chi+i\Gamma/2)$ (see App.\ref{app1}) where $\Gamma=2V^2$ and 
\bea \chi=E_k-E_2+i\gamma_2/2-\f{\Omega_c^2}{4(E_k-E_2+\Delta+i\gamma_3/2)}.\label{trans}
 \eea
 In Fig.\ref{EIT1}  we plot the transmission coefficient $T_k=|\tilde{t}_{k}|^2$ with detuning $(E_k-E_2)$ of the incident probe photon for different values of the control beam Rabi frequency $\Omega_c$. Here we set the loss $\gamma_3$ very small compared to $\Gamma$, i.e., the state $|3\ra$ is metastable. In the absence of the control beam a probe photon is strongly reflected by the emitter due to $|1\ra-|2\ra$ transition, and a Lorentzian dip around $E_k=E_2$ in the transmission line-shape in Fig.\ref{EIT1}(a) reveals it. Thus the 3LE in the absence of a control beam acts as a perfect reflector, and it has been observed in the recent experiments as shown in Fig.2(a) of Refs.\cite{Abdumalikov10,Hoi11}. A narrow transmission window which is much narrower than the Lorentzian dip in the transmission appears at two-photon resonance $E_k-E_2=-\Delta$ as we switch on a weak control beam, $\Omega_c<\Gamma$ in Fig.\ref{EIT1}(b). This induced transparency by the control beam is known as EIT. The EIT is developed due to destructive Fano interference between two allowed atomic transitions which leads to cancellation of the population of the state $|2\ra$, i.e., formation of the `dark state'. As the transition $|1\ra-|2\ra$ gets suppressed at two-photon resonance due to formation of the dark state, the probe photons pass the emitter without being scattered. The width of the transparency window near two-photon resonance increases with an increasing strength of the control beam which is shown in Fig.\ref{EIT1}(c). Finally the ATS appears at a relatively stronger control field, $\Omega_c \ge \Gamma$ (depending on the loss terms) and the splitting between the Autler-Townes doublet is given by the control beam Rabi frequency $\Omega_c$ (see Fig.\ref{EIT1}(d)). The  Autler-Townes doublet forms due to Rabi splitting of the states $|2\ra$ and $|3\ra$.

A ladder-type 3LE (see Fig.\ref{sch}(b)) made of a superconducting qubit was used in two recent experiments \cite{Abdumalikov10,Hoi11} with transmission lines. In these experiments, the lower transition $|1\ra-|2\ra$ of the two allowed transitions of the ladder-type 3LE is coupled to a weak probe beam and the upper transition $|2\ra-|3\ra$ is driven by a control beam. A formula for the transmission amplitude of the probe beam  was derived in Ref.\cite{Abdumalikov10} using the Markovian master equation for the density matrix. We find that their transmission amplitude formula for the driven ladder-type 3LE is exactly similar to our single photon transmission amplitude $\tilde{t}_{k}$ in the driven $\Lambda$-type 3LE when we replace their probe and control beam detunings $\delta \omega_p$ and $\delta \omega_c$ by our $(E_k-E_2)$ and $\Delta$ respectively, and their loss terms $\g_{21}$ and $\g_{31}$ by our $(\g_2+\Gamma)/2$ and $\g_3/2$ respectively. We also identify the probe beam coupling $\Gamma_{21}$ in Ref.\cite{Abdumalikov10} with our $\Gamma$. This similarity is not surprising as the two 3LEs are identical except that the loss rates are practically very different in the two 3LEs. The authors of Refs.\cite{Abdumalikov10,Hoi11} have demonstrated an induced transparency by a strong control beam. However, an EIT transmission line-shape for an arbitrarily weak value of the control beam, $\Omega_c<<\Gamma$ has not been observed in both the experiments. Thus it is not quite clear whether  these experiments demonstrate EIT or they only see the ATS at a strong driving field. A recent theoretical study \cite{Anisimov11} concludes after an objective test of the experimental data that the ATS is preferred to be observed than EIT in these experiments. The state $|3\ra$ is not metastable for a ladder-type 3LE and it has fast pure dephasing or loss, $\g_3>\Gamma$. Therefore it has not been possible to observe a transmission window near two-photon resonance at a weak $\Omega_c$ in the experiments\cite{Abdumalikov10}.

\begin{figure}
\includegraphics[width=8.5cm]{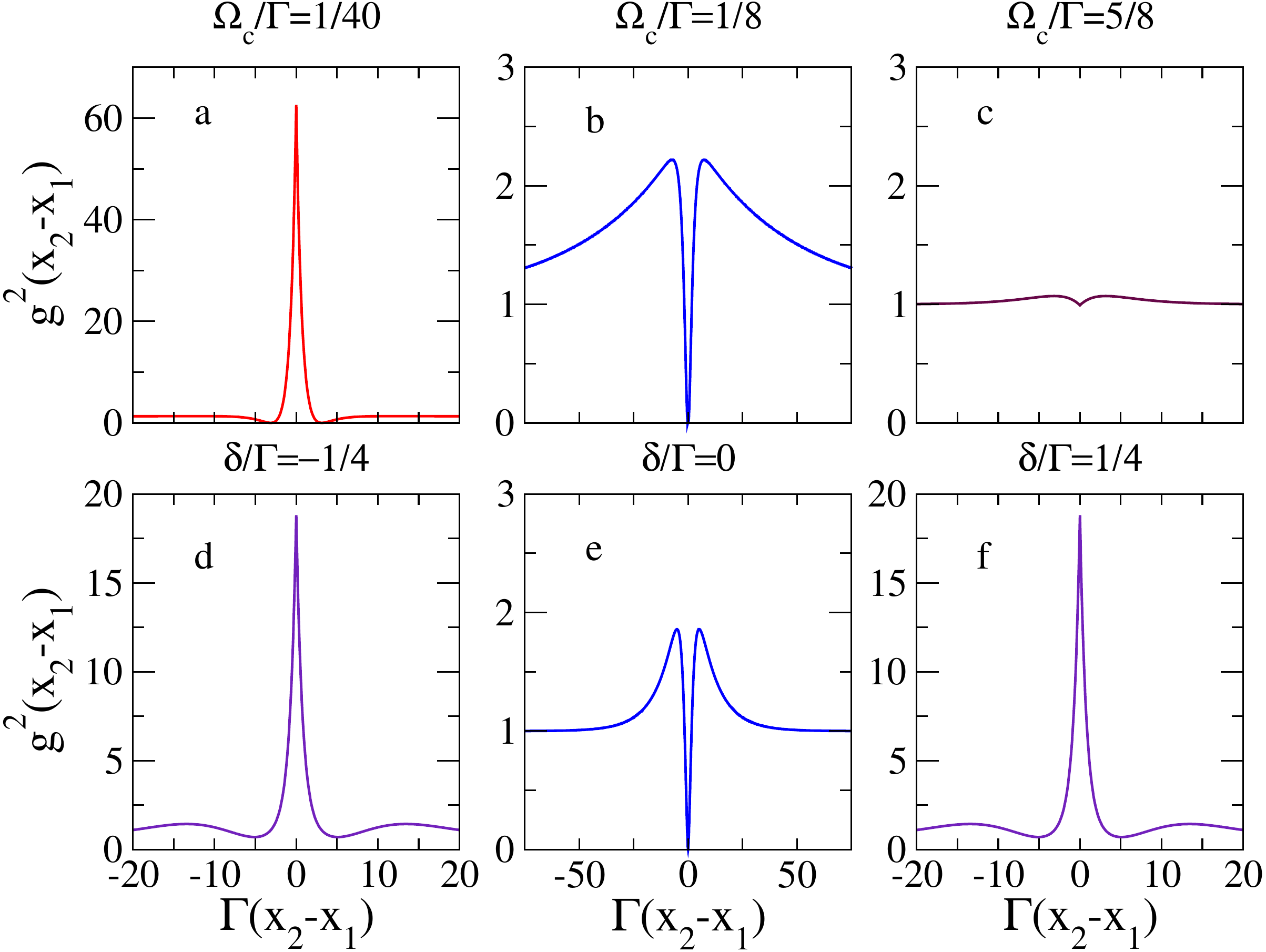}
\caption{Second-order coherence $g^2(x_2-x_1)$ of the transmitted probe photons from a driven $\Lambda$-type emitter at various control beam driving $\Omega_c$ (first row) and two-photon detuning $\delta$ (second row). The probe beam is on two-photon resonance $\delta=(E_k-(E_2-\Delta))=0$ and $\g_3/\Gamma=1/40$ in the first row, and the control beam strength $\Omega_c/\Gamma=3/10$ and $\g_3/\Gamma=1/8$ in the second row. The other parameters are $\Delta=0,~\g_2/\Gamma=0.31$.}
\label{2ndcohT1}
\end{figure}

\section{Second-order coherence}
\label{2OC}
%To calculate the statistics of the scattered probe field from a driven 3LE we need to derive multi-photon scattering state of the probe field. 
A two-photon scattering state of the probe field for a weak control beam was derived in Ref.\cite{Roy11}. However the two-photon state in Ref.\cite{Roy11} is not sufficient to understand how the statistics of scattered field changes with an increasing value of the control beam Rabi frequency, specially at a $\Omega_c$ where the ATS appears. We here derive an exact two-photon scattering state of the probe field for an arbitrary strength of $\Omega_c$ using a method developed recently for an atomic ensemble \cite{Roy13}. We are also able to derive a multi-photon scattering state in the present system (see App.\ref{app1}). This is done in a similar spirit of Ref.\cite{Zheng10}. In the  multi-photon scattering state we consider scattering processes with inelastic exchange of momentum between one pair of photons and elastic exchange of momentum between other all possible pairs. One general way to quantify the statistics is by measuring second-order coherence of the scattered photons. We define second-order coherence by
\bea
g^2(x_2-x_1)=\f{\la \psi |a^{\dg}_{m}(x_1)a^{\dg}_m(x_2)a_m(x_2)a_m(x_1)| \psi \ra}{\la \psi|a^{\dg}_{m}(x_1)a_m(x_1)|\psi \ra \la \psi|a^{\dg}_{m}(x_2)a_m(x_2)|\psi \ra},\nn\\\label{2ndcoh}
\eea
where $m=R$ for the transmitted photons and $m=L$ for the reflected photons for  an incident probe beam from the left. Here $|\psi\ra$ is a $N$-photon scattering Fock state with incident momenta $k_1,k_2..k_N$. A single emitter becomes saturated by a single photon as one emitter can absorb only one photon at a time. Therefore, a strong photon-photon nonlinearity is created by an emitter for two incident photons. %However the relative strength of photon-photon nonlinearity created by an emitter falls with an increasing number of photons of more than two, and  most of the incident photons pass by the emitter by elastic scattering with it. 
Thus we would be able to capture main features of the statistics of scattered photons due to photon-photon nonlinearity and control beam driving by considering  a scattering state of the probe beam with minimum two incident photons. We here find after keeping higher order contributions in the numerator and denominator of Eq.\ref{2ndcoh} (check App.\ref{app2})
\begin{widetext}
\bea
g^2(x_2-x_1)=\f{|\sum_P (t_{k_{P_1}}\pm 1)(t_{k_{P_2}}\pm1)\tilde{h}_{k_{P_1}}(x_1)\tilde{h}_{k_{P_2}}(x_2)+2i\sum_{PQ}V \beta  \,({t}_{k_{P_1}}-1) \Xi_{k_{P_2}}(x_{Q_{12}})\, \tilde{h}_{k_{P_1}}(x_{Q_1})\tilde{h}_{k_{P_2}}(x_{Q_1})\theta(x_{Q_{12}})|^2}{|(t_{k_1}\pm 1)(t_{k_2}\pm1)|^2}%\cos^2((k_1-k_2)(x_2-x_1)/2)}
,\nn\\\label{2ndcoh1}
\eea
\end{widetext}   
where $+$ sign for the transmitted probe beam and $-$ sign for the reflected probe beam. Here $\Xi_{k}(x_1-x_2)=\sum_{j=\pm} d_j \varepsilon_{j}(k)e^{i(s+j\Omega_c/4\beta)|x_1-x_2|}$, $\tilde{h}_k(x)=e^{i k x}\theta(x)/{\sqrt{2}},~\varepsilon_{\pm}(k)=V/(E_k+s\pm \Omega_c/4\beta),~s=-(E_2-\Delta/2)+i(\g_2+\g_3+\G)/4,~\beta=\Omega_c/\sqrt{\epsilon^2+4\Omega_c^2},~d_{\pm}=(1/(2\beta) \pm \epsilon/(2\Omega_c))$ and $\epsilon=-2\Delta+i(\g_2+\G-\g_3)$. We use $P=(P_1,P_2)$ and $Q=(Q_1,Q_2)$ for permutation of $(1,2)$ and $x_{Q_{12}}=x_{Q_1}-x_{Q_2}$. 

Next we discuss nature of second-order coherence at two-photon resonance, i.e., $E_{k_1}=E_{k_2}=E_2-\Delta$. We find from Eq.\ref{trans} that the transmission amplitude $\tilde{t}_k=(t_k+1)/2 \approx 0$ when $\Omega_c \sim 0$ and $\tilde{t}_k \approx 1$ when $\Omega_c \sim \Gamma$. Here we assume that $\g_2<\Gamma$ for both $\Lambda$- and ladder-type 3LE. In the absence of the control beam the 3LE-waveguide system reduces to a two-level emitter coupled to a probe beam, and we find bunching of the transmitted photons due to the inelastic two-photon bound state (the second term of the numerator in Eq.\ref{2ndcoh1}) when $E_k=E_2$. It has been demonstrated in a recent experiment \cite{Hoi12} with a two-level emitter. When $T_k=1$ in the presence of a strong control beam driving, the photon-photon correlation due to the inelastic two-photon bound state becomes negligible, and the second-order coherence of the transmitted probe beam is mostly determined by the first term of the numerator in Eq.\ref{2ndcoh1}. Then the numerator and denominator of $g^2(x_2-x_1)$ become the same, and we have $g^2(x_2-x_1)=1$. At an intermediate control beam driving, $\Omega_{c0}^2=\gamma_3(\Gamma-\g_2)$, $t_k$ vanishes, and the single probe photon transmission amplitude $\tilde{t}_k=1/2$. At $\Omega_{c0}$ the numerator in Eq.\ref{2ndcoh1} vanishes at $x_1=x_2$ and the numerator is non-zero when $x_1 \ne x_2$. Therefore $g^2(x_2-x_1)$ exhibits antibunching of the transmitted probe beam at two-photon resonance when $\Omega_c=\Omega_{c0}$ \cite{Zheng12}. Physically the antibunching occurs due to interference between the partially transmitted probe photons and the ineleastic two-photon bound state.
\begin{figure}
\includegraphics[width=8.5cm]{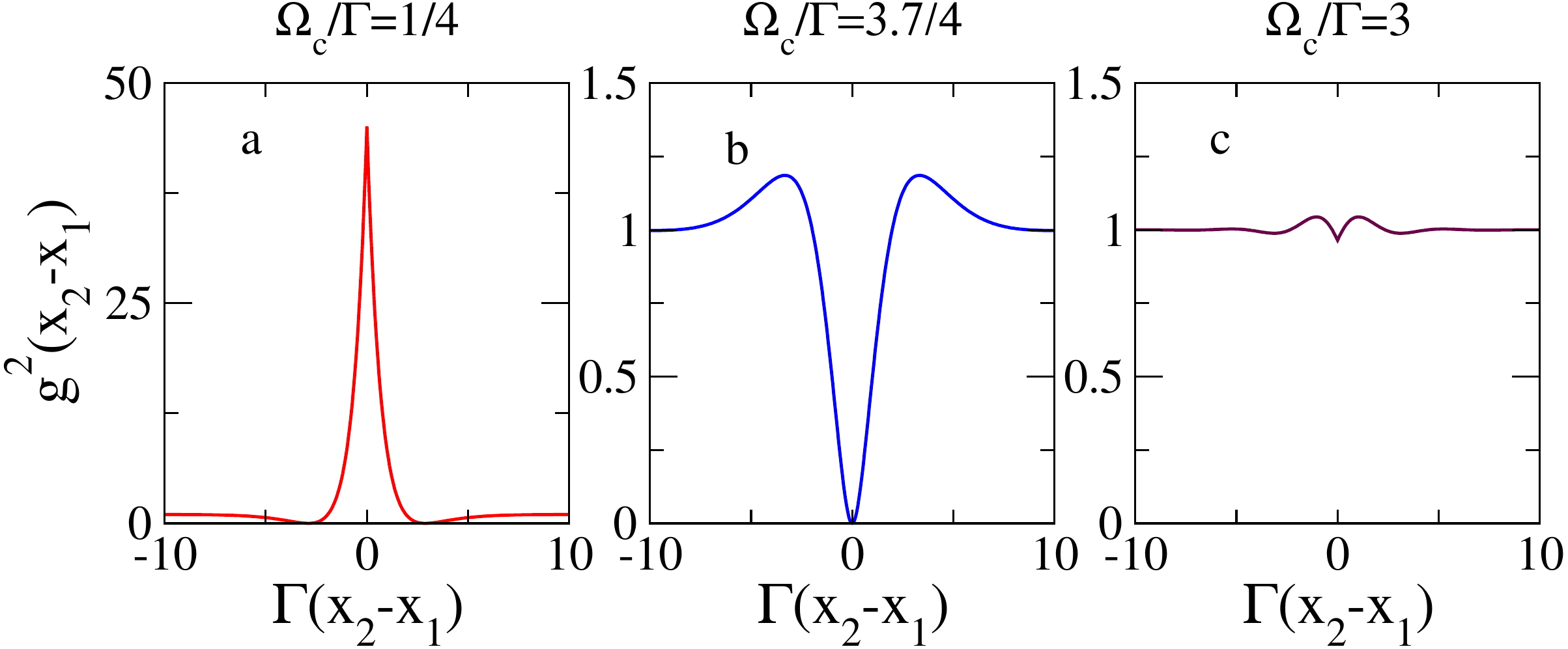}
\caption{Second-order coherence $g^2(x_2-x_1)$ of the transmitted probe beam from a driven ladder-type emitter at various control beam driving $\Omega_c$. The incident probe beam is a coherent state wave-packet with $k_0=E_2-\Delta>>\Delta_k=\Gamma/40,~\bar{n}=1$. The other parameters are $\Delta=0$, $\Gamma/2\pi$=11 MHz, $\gamma_{2}/2\pi$=3.4 MHz and $\gamma_{3}/2\pi$=13.8 MHz.}
\label{2ndcohT2}
\end{figure}

We show the above discussed behavior of $g^2(x_2-x_1)$ of the transmitted probe photons for a driven $\Lambda$-type 3LE in Fig.\ref{2ndcohT1} where we set the loss term $\g_3$ to be very small compared to $\Gamma$, such that the state $|3\ra$ is metastable. We show bunching of transmitted probe photons in Fig.\ref{2ndcohT1}(a) for a very weak control beam when $T_k \approx 0$. We kept the incident probe beam on two-photon resonance, $E_{k_1}=E_{k_2}=E_2-\Delta$. Next we slowly increase the strength of Rabi frequency of the control beam. We find from Fig.\ref{2ndcohT1}(b) that $g^2(x_2-x_1)$ shows antibunching of the transmitted probe photons when $\Omega_c$ is near $\sqrt{\g_3(\Gamma-\g_2)}$. The antibunching implies that two probe photons can not transmit through the emitter simultaneously. This happens for a Rabi frequency when a complete dark state is not yet formed. As we further increase $\Omega_c$ a dark state is formed and $T_k$ becomes unity at the two-photon resonance. There $g^2(x_2-x_1)=1$ as shown in Fig.\ref{2ndcohT1}(c), and the incident probe photons are not scattered by the driven emitter. When frequency of the incident probe beam is detuned from the two-photon resonance condition of EIT, $g^2(x_2-x_1)$ shows bunching (check Figs.\ref{2ndcohT1}(d,f)) as one photon then gets strongly scattered by the emitter. 

Finally we discuss second-order coherence of the transmitted probe beam when the incident probe beam is a coherent state wave-packet. The incident coherent state wave-packet is given by $|\alpha\ra=e^{a^{\dg}_{\alpha}-\bar{n}/2}|\varphi\ra$ where $a^{\dg}_{\alpha}=\int dk \alpha(k)a^{\dg}(k)$, $|\varphi\ra$ is vacuum state and the mean photon number is $\bar{n}=\int dk |\alpha(k)|^2$. Here $a^{\dg}(k)=\int dx~ e^{ikx} a^{\dg}_R(x)/\sqrt{2\pi}$ for an incident wave-packet from the left. We consider the mean photon number of the coherent state wave-packet $\bar{n} \le 1$ and choose a Gaussian wave-packet \cite{Zheng10}
\bea
\alpha(k)=\f{\sqrt{\bar{n}}}{(2\pi\Delta_k^2)^{1/4}}{\rm exp}\big(-\f{(k-k_0)^2}{4\Delta_k^2}\big),
\eea 
where $\Delta_k$ is the width of the wave-packet and $k_0$ is mean momentum (or energy) of the wave-packet. We choose $k_0=E_2-\Delta>>\Delta_k=\Gamma/40$. The statistics of scattered probe photons for a coherent state wave-packet input remains similar to that of a Fock state input, provided that the bandwidth of the coherent state input is significantly narrower than the emitter's line-width. We show this in Fig.\ref{2ndcohT2} for a ladder-type 3LE with a large dephasing loss from the state $|3\ra$ \cite{Abdumalikov10}. The nature of second-order coherence changes from bunching (Fig.\ref{2ndcohT2}(a)) to antibunching (Fig.\ref{2ndcohT2}(b)) to one for coherent state (Fig.\ref{2ndcohT2}(c)) as $\Omega_c$ is increased from a weak to a strong value.

\section{Conclusion}
\label{conl}
In conclusion, we have shown that second-order coherence of the scattered probe  photons from a $\Lambda$- or ladder-type 3LE can be tuned by changing Rabi frequency of the control beam. The single-photon transmission coefficient of the probe beam from a ladder-type 3LE at different strength of the control beam has been already measured in recent experiments \cite{Abdumalikov10, Hoi11}. The second-order coherence in a $\Lambda$- or ladder-type 3LE can be measured experimentally using a Hanbury-Brown-Twiss measurement setup \cite{Hoi12}. One important generalization of our present calculation would be to derive multi-photon scattering states and second-order coherence of the scattered probe photons from a quantum nonlinear medium consisting of multiple interacting multi-level emitters \cite{Deutsch92, Firstenberg13}. 

\section{Acknowledgments}
D.R. gratefully acknowledges the support of the U.S. Department of Energy through LANL/LDRD Program for this work.

\appendix
\section{Scattering states}
\label{app1}
The multi-photon scattering states of the probe beam are calculated in the even-odd basis of probe photons, defined by $a_{e}(x)=(a_{R}(x)+a_{L}(-x))/\sqrt{2}$ and $a_{o}(x)=(a_{R}(x)-a_{L}(-x))/\sqrt{2}$. In the even-odd basis the Hamiltonian is $\mathcal{H}=\mathcal{H}_e+\mathcal{H}_o,$  where
\bea
\mathcal{H}_e&=&-i v_g\int dx~ a^\dg_{e}(x)\partial_xa_{e}(x) + \mathcal{H}_{3LE} \nn\\&+&\bar{V} \big(a^\dg_{e}(0)|1\ra\la2|+|2\ra\la1|a_{e}(0)\big),~{\rm and}~\nn\\
 \mathcal{H}_o &=&-iv_g \int dx~ a^\dg_{o}(x)\partial_ xa_{o}(x), \nn
\eea
where $\bar{V}=\sqrt{2}V$. In the transformed basis, the even part $\mathcal{H}_e$  contains interaction of the even field modes with the 3LE while the odd part $\mathcal{H}_o$ is only the kinetic energy of the noninteracting odd field modes. Therefore, nontrivial scattering of probe photons from the 3LE occurs only in the even sector of the Hamiltonian. In the following we give explicit results for the scattering of even photon modes.

%\appendix
\subsection{Single-photon scattering state}
An exact single probe photon scattering state has been derived previously \cite{Roy11}. However we here state the main results in the single photon sector for the shake of completeness. The single-photon eigenstate of the full system is 
\bea
|k\ra&=&\int dx \{A_1[g_k(x)a^{\dg}_e(x)|0,1\ra+ e_k|0,2\ra +f_k|0,3\ra]\nn\\&+&B_1h_k(x)a^{\dg}_{o}(x)|0,1\ra\},\label{sinstate}
\eea
where the constants $A_1$ and $B_1$ keep track of the non-equilibrium open boundary condition, i.e., the incident photon is coming from which side of the emitter. For a photon coming from the left (right-moving photon), the incoming state is $|k\ra_{\text in}=(1/\sqrt{2\pi})\int dx e^{ikx}a^{\dg}_R(x)|0,1\ra$, and we find $A_1=B_1=1/\sqrt{2}$. Here $g_k(x)$ and $h_k(x)$ are the amplitude of a single photon in the even and odd field modes when the emitter in the ground state. For an incident photon coming from the left, $g_k(x<0)=h_k(x<0)=e^{ikx}/\sqrt{2\pi}$. The amplitude of the excited states $|2\ra$ and $|3\ra$ are respectively given by $e_k$ and $f_k$. The basis state $|0,i\ra$ denotes zero photon in the waveguide and the emitter in the $i^{\text{th}}$ state.  We find different amplitudes in the state in Eq.(\ref{sinstate}) by solving a set of coupled linear differential equations obtained from the stationary Schr{\"o}dinger equation, $\mathcal{H}|k\ra=E_k|k\ra$ with $E_k=v_gk$.
\bea
-iv_g\partial_xg_{k}(x)-E_k g_{k}(x)+\bar{V}e_k\delta(x)&=&0,\nn \\
(E_2-i\gamma_2/2-E_k)e_k+\bar{V}g_k(x)\delta(x)+\f{\Omega_c}{2}f_k&=&0,\nn \\
(E_2-\Delta-i\gamma_3/2-E_k)f_k+\f{\Omega_c}{2}e_k&=&0,\nn \\
-iv_g\partial_xh_{k}(x)-E_k h_{k}(x)&=&0.
\eea 
We use a regularization of the following form for the amplitudes across the emitter position, $g_k(0)=[g_k(0-)+g_k(0+)]/2$ and the initial boundary conditions to solve the above differential equations of the amplitudes. We find $g_k(x)=h_k(x)\Big[\theta(-x)+t_k\theta(x)\Big],~h_k(x)=e^{ikx}/\sqrt{2\pi},~e_k=\bar{V}/(\sqrt{2\pi}(\chi+i\Gamma/2)),~f_k=0.5\Omega_ce_k/(E_k-E_2+\Delta+i\gamma_3/2)$ and $t_k=(\chi-i\Gamma/2)/(\chi+i\Gamma/2)$. Here $\theta(x)$ is the step function, $\Gamma=\bar{V}^2/v_g=2V^2/v_g$ and
\bea
\chi=E_k-E_2+i\gamma_2/2-\f{\Omega_c^2}{4(E_k-E_2+\Delta+i\gamma_3/2)}~.
\eea 
Now onwards we set $v_g=1$. For an incident photon from the left,  we find the single-photon transmission and reflection amplitudes of the original right and left moving photons are given by $\tilde{t}_{k}=(1+t_k)/2=\chi/(\chi+i\Gamma/2)$ and $\tilde{r}_k=(t_k-1)/2=-0.5i\Gamma/(\chi+i\Gamma/2)$. 

\subsection{Two-photon scattering state}
A two-photon scattering state of the probe beam in the present system has been first derived in Ref.\cite{Roy11} by one of us (D.R.). However it was calculated only for a weak control beam and the ATS in this system appears only at a strong driving by the control beam. Here we derive an exact two-photon scattering state for an arbitrary strength of the control beam. We consider both the incident probe photons from the left side of the emitter (right-moving). Thus an incoming two-photon state is given by
\bea
|k_1,k_2\ra_{in}=\int dx_1 dx_2\phi_{\bf k}(x_1,x_2)\f{1}{\sqrt{2}}a^{\dg}_R(x_1)a^{\dg}_R(x_2)|0,1\ra, \nn\\\label{instate}
\eea
where $\phi_{\bf k}(x_1,x_2)=(e^{ik_1x_1+ik_2x_2}+e^{ik_1x_2+ik_2x_1})/(2\sqrt{2}\pi)$ with ${\bf k}=(k_1,k_2)$. We decompose the two-photon incoming state into $ee,~oo$, and $eo$ subspaces, and determine the full scattering eigenstates in the different subspaces separately. The two- or multi-photon transport in the 1D waveguide is strongly correlated since the emitter can absorb only one photon at a time, and the second photon in the probe beam has to wait a time order of $\Gamma^{-1}$ for the first photon to be spontaneously re-emitted by the emitter. By the process of absorption and spontaneous emission the emitter mediates strong interactions between photons. However, there can be another process where the second photon coming within the time interval of $\Gamma^{-1}$ of the first-photon absorption by the emitter can excite the emitter for stimulated emission of the first photon into the state of the second photon. This process would develop a bound state of the two photons, and we here calculate the two-photon bound-state contribution exactly in our two-photon scattering state. The general two-photon scattering state in our system is of the form:
\begin{widetext}
\bea
&&|k_1,k_2\ra=\int dx_1dx_2\Big[A_2\big\{g(x_1,x_2)\f{1}{\sqrt{2}}a^{\dg}_e(x_1)a^{\dg}_e(x_2)|0,1\ra+(e(x_1)a^{\dg}_e(x_1)|0,2\ra+f(x_1)a^{\dg}_e(x_1)|0,3\ra)\delta(x_2)\big\}\label{wavefn}\\&&+B_2\big\{j(x_1;x_2)a^{\dg}_e(x_1)a^{\dg}_o(x_2)|0,1\ra+(v(x_1)a^{\dg}_o(x_1)|0,2\ra+w(x_1)a^{\dg}_o(x_1)|0,3\ra)\delta(x_2)\big\}+C_2h(x_1,x_2)\f{1}{\sqrt{2}}a^{\dg}_o(x_1)a^{\dg}_o(x_2)|0,1\ra\Big], \nn
\eea 
\end{widetext}
where two-photon amplitudes in the $ee$ and $oo$ subspaces satisfy $g(x_1,x_2)=g(x_2,x_1)$ and $h(x_1,x_2)=h(x_2,x_1)$ due to Bose statistics of photons. Here $e(x),f(x)~(v(x),w(x))$ are the amplitude of one photon in the $e~(o)$-subspace while the 3LE in the excited state $|2\ra$ and $|3\ra$ respectively. Here $A_2,B_2$ and $C_2$ identify the boundary condition for the incoming photons. For two incident photons from the left (right-moving photons), $A_2=B_2=C_2=1/2$. Note that we here express the two-photon scattering state in the space of free probe photons and the 3LE. This is similar to the single photon scattering states in Eq.(\ref{sinstate}). We obtain the following coupled linear differential equations by using the stationary Schr{\"o}dinger equation $\mathcal{H}|k_1,k_2\ra=E_{\bf k}|k_1,k_2\ra$ for the two-photon scattering state in Eq.(\ref{wavefn}) where $E_{\bf k}=E_{k_1}+E_{k_2}$.
\begin{widetext}
\bea
\Big(-i(\partial_{x_1}+\partial_{x_2})-E_{\bf k}\Big)g(x_1,x_2)+\f{\bar{V}}{\sqrt{2}}[e(x_1)\delta(x_2)+\delta(x_1)e(x_2)]&=&0, \label{3lese1} \\
\Big(-i\partial_x-E_{\bf k}+E_2-i\f{\gamma_2}{2}\Big)e(x)+\f{\bar{V}}{\sqrt{2}}[g(0,x)+g(x,0)]+\f{\Omega_c}{2}f(x)&=&0, \label{3lese2} \\
\Big(-i\partial_x-E_{\bf k}+E_2-\Delta-i\f{\gamma_3}{2}\Big)f(x)+\f{\Omega_c}{2}e(x)&=&0, \label{3lese3} \\
\Big(-i(\partial_{x_1}+\partial_{x_2})-E_{\bf k}\Big)j(x_1;x_2)+\bar{V}\delta(x_1)v(x_2)&=&0,\label{3lese4}\\
\Big(-i\partial_x-E_{\bf k}+E_2-i\f{\gamma_2}{2}\Big)v(x)+\bar{V}j(0;x)+\f{\Omega_c}{2}w(x)&=&0,\label{3lese5}  \\
\Big(-i\partial_x-E_{\bf k}+E_2-\Delta-i\f{\gamma_3}{2}\Big)w(x)+\f{\Omega_c}{2}v(x)&=&0, \label{3lese6} \\
\Big(-i(\partial_{x_1}+\partial_{x_2})-E_{\bf k}\Big)h(x_1,x_2)&=&0.\label{3lese7}
\eea
We find from Eq.(\ref{3lese1}), $g(0+,x)=g(0-,x)-\f{i\bar{V}}{\sqrt{2}}e(x)$. To solve above set of coupled differential equations we rewrite them in a matrix notation \cite{Roy13}. Thus we find from Eqs.(\ref{3lese2},\ref{3lese3})
\bea
\partial_x\left( \begin{array}{c}  e(x) \\ f(x) \end{array}\right)=i\left( \begin{array}{cc} E_{\bf k}-E_2+i(\bar{V}^2+\gamma_2)/2  & -\Omega_c/2 \\ -\Omega_c/2 & E_{\bf k}-E_2+\Delta+i\gamma_3/2  \end{array}\right)\left( \begin{array}{c}  e(x) \\ f(x) \end{array}\right)-\sqrt{2}i\bar{V}g(x,0-)\left( \begin{array}{c}  1 \\ 0 \end{array}\right).\label{smat}
\eea
We call the $2\times 2$ square matrix in Eq.(\ref{smat}) by $\overleftrightarrow{\bf A}$. Here we define $\bar{V}^2+\gamma_2=\tilde{\gamma}_2,~-2\Delta +i(\tilde{\gamma}_2-\gamma_3)=\epsilon$. The eigenvalues and eigenvectors of the above square matrix  are $\lambda_{\mp}=E_{\bf k}+s \mp t$ with $s=-E_2+\Delta/2+i(\tilde{\gamma}_2+\gamma_3)/4$, $t= \sqrt{\epsilon^2+4\Omega_c^2}/4$ and
\bea
|\lambda_-\ra= \left( \begin{array}{c}  \f{-\epsilon+\sqrt{4\Omega_c^2+\epsilon^2}}{2\Omega_c} \\ 1 \end{array}\right),~~|\lambda_+\ra= \left( \begin{array}{c}  \f{-\epsilon-\sqrt{4\Omega_c^2+\epsilon^2}}{2\Omega_c} \\ 1 \end{array}\right).
\eea
We form a $2\times2$ square matrix $\overleftrightarrow{\bf P}$ using the eigenvectors of $\overleftrightarrow{\bf A}$, thus $\overleftrightarrow{\bf P}=(|\lambda_-\ra, |\lambda_+\ra)$, and the inverse of $\overleftrightarrow{\bf P}$ is given by
\bea
\overleftrightarrow{\bf P}^{-1}=\left( \begin{array}{cc} \f{\Omega_c}{\sqrt{4\Omega_c^2+\epsilon^2}}  & \f{\epsilon+\sqrt{4\Omega_c^2+\epsilon^2}}{2\sqrt{4\Omega_c^2+\epsilon^2}} \\ -\f{\Omega_c}{\sqrt{4\Omega_c^2+\epsilon^2}}  & \f{-\epsilon+\sqrt{4\Omega_c^2+\epsilon^2}}{2\sqrt{4\Omega_c^2+\epsilon^2}} \end{array}\right),~~{\rm so~that}~~\overleftrightarrow{\bf P}^{-1}\overleftrightarrow{\bf A}\overleftrightarrow{\bf P}=\left( \begin{array}{cc} \lambda_-  & 0 \\ 0 &  \lambda_+ \end{array}\right).
\eea
Therefore we can now write,
\bea
\partial_x\left( \begin{array}{c}  \tilde{e}(x) \\ \tilde{f}(x) \end{array}\right)&=&i\left( \begin{array}{cc} \lambda_-  & 0 \\ 0 &  \lambda_+ \end{array}\right)\left( \begin{array}{c}  \tilde{e}(x) \\ \tilde{f}(x) \end{array}\right)-\f{\Omega_c \sqrt{2} i\bar{V} g(x,0-)}{\sqrt{4\Omega_c^2+\epsilon^2}}\left( \begin{array}{c} 1 \\  -1 \end{array}\right),\label{3lerotex}
\eea
where $\tilde{e}(x)=\f{\Omega_c}{\sqrt{4\Omega_c^2+\epsilon^2}}e(x)+\f{\epsilon+\sqrt{4\Omega_c^2+\epsilon^2}}{2\sqrt{4\Omega_c^2+\epsilon^2}}f(x)~,~\tilde{f}(x)=-\f{\Omega_c}{\sqrt{4\Omega_c^2+\epsilon^2}}e(x)+\f{-\epsilon+\sqrt{4\Omega_c^2+\epsilon^2}}{2\sqrt{4\Omega_c^2+\epsilon^2}}f(x)$. We can find the original amplitudes $e(x),f(x)$ after determining $\tilde{e}(x), \tilde{f}(x)$. The inverse transformations are $f(x)=\tilde{e}(x)+\tilde{f}(x)$ and $e(x)=\f{\sqrt{4\Omega_c^2+\epsilon^2}-\epsilon}{2\Omega_c}\tilde{e}(x)-\f{\sqrt{4\Omega_c^2+\epsilon^2}+\epsilon}{2\Omega_c}\tilde{f}(x)$. We have continuity relations, $\tilde{e}(0-)=\tilde{e}(0+),~\tilde{f}(0-)=\tilde{f}(0+)$.

Now we find $\tilde{e}(x)$ and $\tilde{f}(x)$ by solving the two equations in Eq.(\ref{3lerotex}). We derive at $x<0$ 
 \bea
\tilde{e}(x)&=&\f{\beta}{2\pi}(\varepsilon_{k_1}e^{ik_2x}+\varepsilon_{k_2}e^{ik_1x}),~~{\rm where}~~\varepsilon_{k}=\f{\bar{V}}{E_k+s-\Omega_c/(4\beta)},~~\beta=\f{\Omega_c}{\sqrt{4\Omega_c^2+\epsilon^2}},\nn\\
\tilde{f}(x)&=&-\f{\beta}{2\pi}(\varsigma_{k_1}e^{ik_2x}+\varsigma_{k_2}e^{ik_1x}),~~{\rm where}~~\varsigma_{k}=\f{\bar{V}}{E_k+s+\Omega_c/(4\beta)}.\nn
\eea
Thus we get using the above relations at $x<0$, 
\bea
e(x)&=&\f{1}{4\pi}\big[(\varepsilon_{k_1}+\varsigma_{k_1})e^{ik_2x}+\f{\epsilon\beta}{\Omega_c}(\varsigma_{k_1}-\varepsilon_{k_1})e^{ik_2x}\big]+(k_1\leftrightarrow k_2),\nn\\
f(x)&=&\f{\beta}{2\pi}((\varepsilon_{k_2}-\varsigma_{k_2})e^{ik_1x}+(\varepsilon_{k_1}-\varsigma_{k_1})e^{ik_2x}),
\eea
which we employ for finding $g(x_1,x_2)$ at $x_1>0,x_2<0$ from the relation after Eq.(\ref{3lese7})
\bea
g(x_1,x_2)=\f{1}{2\sqrt{2}\pi}(t_{k_1}e^{ik_1x_1+ik_2x_2}+t_{k_2}e^{ik_2x_1+ik_1x_2})~~{\rm where}~~t_{k}=1-\f{i\bar{V}}{2}(\varepsilon_{k}+\varsigma_{k})+\f{i\bar{V}\epsilon\beta}{2\Omega_c}(\varepsilon_k-\varsigma_k).
\eea
Here we remind that $t_k$ in the above relation is the same to $t_k$ which appear in our single photon scattering state, $g_k(x)$ in Eq.(\ref{sinstate}). Next we evaluate $\tilde{e}(x)$ and $\tilde{f}(x)$ at $x>0$ using the Eq.(\ref{3lerotex}),
\bea
\tilde{e}(x)&=&c_ee^{i\lambda_-x}+\f{\beta}{2\pi}(t_{k_1}\varepsilon_{k_2}e^{ik_1x}+t_{k_2}\varepsilon_{k_1}e^{ik_2x}),~{\rm where}~~c_e=\f{i\bar{V}\beta}{2\pi}\big((1-\f{\epsilon \beta}{\Omega_c})\varepsilon_{k_1} \varepsilon_{k_2}+\f{1}{2}(1+\f{\epsilon \beta}{\Omega_c})(\varepsilon_{k_1} \varsigma_{k_2}+ \varepsilon_{k_2} \varsigma_{k_1})\big), \nn \\
\tilde{f}(x)&=&c_fe^{i\lambda_+x}-\f{\beta}{2\pi}(t_{k_1}\varsigma_{k_2}e^{ik_1x}+t_{k_2}\varsigma_{k_1}e^{ik_2x}),~{\rm where}~~c_f=\f{i\bar{V}\beta}{2\pi}\big(-(1+\f{\epsilon \beta}{\Omega_c})\varsigma_{k_1} \varsigma_{k_2}+\f{1}{2}(-1+\f{\epsilon \beta}{\Omega_c})(\varepsilon_{k_1} \varsigma_{k_2}+ \varepsilon_{k_2} \varsigma_{k_1})\big).\nn \eea
Now we can write the form of $e(x>0),f(x>0)$ from the above expressions.
\bea
e(x)&=&(\f{1}{2\beta}-\f{\epsilon}{2\Omega_c})c_ee^{i\lambda_-x}-(\f{1}{2\beta}+\f{\epsilon}{2\Omega_c})c_fe^{i\lambda_+x}+\big[\f{1}{4\pi}(\varepsilon_{k_2}+\varsigma_{k_2})t_{k_1}e^{ik_1x}+\f{\epsilon\beta}{4\pi\Omega_c}(\varsigma_{k_2}-\varepsilon_{k_2})t_{k_1}e^{ik_1x}+(k_1\leftrightarrow k_2)\big],\nn\\
f(x)&=&\tilde{e}(x)+\tilde{f}(x)=c_ee^{i\lambda_-x}+c_fe^{i\lambda_+x}+\f{\beta}{2\pi}(t_{k_1}(\varepsilon_{k_2}-\varsigma_{k_2})e^{ik_1x}+t_{k_2}{(\varepsilon_{k_1}-\varsigma_{k_1})e^{ik_2x})}.
\eea
Using the last results we find the two-photon amplitude in the $ee$ subspace at $x_1>x_2>0$,
\bea
g(x_1,x_2)&=&\f{1}{2\sqrt{2}\pi}(t_{k_1}t_{k_2}e^{ik_1x_1+ik_2x_2}+t_{k_1}t_{k_2}e^{ik_2x_1+ik_1x_2})-\f{i\bar{V}}{\sqrt{2}}\big[(\f{1}{2\beta}-\f{\epsilon}{2\Omega_c})c_ee^{i\lambda_-x_1+i(E_{\bf k}-\lambda_-)x_2}\nn\\&-&(\f{1}{2\beta}+\f{\epsilon}{2\Omega_c})c_fe^{i\lambda_+x_1+i(E_{\bf k}-\lambda_+)x_2}\big].\nn
\eea
Following the above procedure we can derive the amplitudes in the $eo$ and $oo$ subspaces. The amplitudes of a two-photon scattering state of the probe beam for general values of $x_1$ and $x_2$ can be expressed in a compact notation as bellow,
\bea
g(x_1,x_2)&=&\f{1}{\sqrt{2!}} \Big[\sum_P g_{k_{P_1}}(x_1)g_{k_{P_2}}(x_2)+\sum_{PQ}B^{(2)}_{k_{P_1},k_{P_2}} (x_{Q_1},x_{Q_2})\theta(x_{Q_2}) \Big],\label{2phs} \\
e(x)&=& \f{\beta}{\sqrt{2 \pi}}\sum_P g_{k_{P_1}}(x)\Gamma_{k_{P_2}}(0) +\beta \sum_P (1-t_{k_{P_1}})\Gamma_{k_{P_2}}(x) h_{k_{P_1}}(x)h_{k_{P_2}}(x){ \theta(x) }, \\
f(x)&=&\f{\beta}{\sqrt{2 \pi}}\sum_P g_{k_{P_1}}(x)\Upsilon_{k_{P_2}}(0) +\beta \sum_P (1-t_{k_{P_1}})\Upsilon_{k_{P_2}}(x) h_{k_{P_1}}(x)h_{k_{P_2}}(x) { \theta(x)}, \\
j(x_1;x_2)&=& \sum_P {g_{k_{P_1}}(x_1)h_{k_{P_2}}(x_2)},~~v(x)= \f{\beta}{\sqrt{ 2 \pi}}\sum_P h_{k_{P_1}}(x)\Gamma_{k_{P_2}}(0),~~w(x)=\f{\beta}{\sqrt{ 2 \pi}}\sum_P h_{k_{P_1}}(x)\Upsilon_{k_{P_2}}(0), \\
h(x_1,x_2)&=& \f{1}{\sqrt{2!}}  \sum_P {h_{k_{P_1}}(x_1)h_{k_{P_2}}(x_2)},~~{\rm where}~~
\Gamma_{k}(x-y)= \big( d_- \varepsilon_{k}e^{i(s-t)|x-y|}+d_+ \varsigma_{k}e^{i(s+t)|x-y|} \big), d_{\pm}=(\f{1}{2\beta}\pm \f{\epsilon}{2\Omega_c}), \\
\Upsilon_{k}(x-y)&=& \big(  \varepsilon_{k}e^{i(s-t)|x-y|}-\varsigma_{k}e^{i(s+t)|x-y|} \big),~B^{(2)}_{k_{P_1},k_{P_2}} (x_{Q_1},x_{Q_2})=-i \bar{V} \beta  \, (1-t_{k_{P_1}}) \Gamma_{k_{P_2}}(x_{Q_{12}})\, h_{k_{P_1}}(x_{Q_1})h_{k_{P_2}}(x_{Q_1})\theta(x_{Q_{12}}).\nn
\label{notation}
\eea
Here $P=(P_1,P_2)$ and $Q=(Q_1,Q_2)$ are permutation of $(1,2)$, and $x_{Q_{12}}=x_{Q_1}-x_{Q_2}$. The expression $B^{(2)}_{k_{P_1},k_{P_2}}(x_{Q_1},x_{Q_2})$ in Eq.\ref{2phs} is the contribution coming from the two-photon bound state.

\subsection{Three-photon scattering state}
%A multi-photon scattering state of the probe beam in this system has not been derived before even for a weak control beam. 
 Now we calculate a three-photon scattering state in the present system for an arbitrary control beam. In the three-photon scattering state, we consider scattering processes with inelastic exchange of momentum between any one pair of photons and elastic exchange of momentum between all other possible pairs. %This procedure can be extended for four or more photons. 
 We write a three-photon scattering state as following
\bea
&&|k_1,k_2,k_3\ra=\int dx_1dx_2 dx_3\Big[A_3\big\{g(x_1,x_2,x_3)\f{1}{\sqrt{3!}}a^{\dg}_e(x_1)a^{\dg}_e(x_2)a^{\dg}_e(x_3)|0,1\ra+e_2(x_1,x_2)\delta(x_3)a^{\dg}_e(x_1)a^{\dg}_e(x_2)|0,2\ra \nn \\ &&+e_3(x_1,x_2)\delta(x_3)a^{\dg}_e(x_1)a^{\dg}_e(x_2)|0,3\ra\big\}+B_3\big\{j(x_1,x_2;x_3)a^{\dg}_e(x_1)a^{\dg}_e(x_2)a^{\dg}_o(x_3)|0,1\ra+v_2(x_1;x_3)\delta(x_2)a^{\dg}_e(x_1)a^{\dg}_o(x_3)|0,2\ra \nn \\
&&+v_3(x_1;x_3)\delta(x_2)a^{\dg}_e(x_1)a^{\dg}_o(x_3)|0,3\ra\big\}+C_3\big\{f(x_1;x_2,x_3)a^{\dg}_e(x_1)a^{\dg}_o(x_2)a^{\dg}_o(x_3)|0,1\ra+w_2(x_2,x_3)\delta(x_1)a^{\dg}_o(x_2)a^{\dg}_o(x_3)|0,2\ra \nn \\
&&+w_3(x_2,x_3)\delta(x_1)a^{\dg}_o(x_2)a^{\dg}_o(x_3)|0,3\ra\big\}+D_3~h(x_1,x_2,x_3)\f{1}{\sqrt{3!}}a^{\dg}_o(x_1)a^{\dg}_o(x_2)a^{\dg}_o(x_3)|0,1\ra\Big],
\label{wavefn3}
\eea 
where $A_3=D_3=1/2^{3/2}$ and $B_3=C_3=3/2^{3/2}$. We obtain the following set of coupled linear differential equations by substituting the above ansatz for the three-photon scattering state in the three-photon stationary Schr{\"o}dinger equation with total energy of the three photons $E_{\bf k}=E_{k_1}+E_{k_2}+E_{k_3}$. The equations in the even photon field basis are
\bea
&&\Big(-i(\partial_{x_1}+\partial_{x_2}+\partial_{x_3})-E_{\bf k}\Big)g(x_1,x_2,x_3)+\sqrt{\f{2}{3}}\,\bar{V}[e_2(x_1,x_2)\delta(x_3)+e_2(x_1,x_3)\delta(x_2)+e_2(x_2,x_3)\delta(x_1)]=0, \label{3SE1} \\
&&\Big(-i(\partial_{x_1}+\partial_{x_2})-E_{\bf k}+E_2-i\f{\gamma_2}{2}\Big)e_2(x_1,x_2)+\f{\bar{V}}{\sqrt{3!}}[g(x_1,x_2,0)+g(x_1,0,x_2)+g(0,x_1,x_2)]+\f{\Omega_c}{2}e_3(x_1,x_2)=0, \label{3SE2} \\
&&\Big(-i(\partial_{x_1}+\partial_{x_2})-E_{\bf k}+E_2-\Delta-i\f{\gamma_3}{2}\Big)e_3(x_1,x_2)+\f{\Omega_c}{2}e_2(x_1,x_2)=0. \label{3SE3} 
\eea
We find from Eq.\ref{3SE1}, $g(0+,x_2,x_3)=g(0-,x_2,x_3)-i\bar{V}\sqrt{\f{2}{3}} \,e_2(x_2,x_3)$. To solve the above set of coupled differential equations we rewrite them in a matrix notation as before in the two-photon scattering state; we find from Eqs.\ref{3SE2},\ref{3SE3}
\bea
(\partial_{x_1}+\partial_{x_2})\left( \begin{array}{c}  e_2(x_1,x_2) \\ e_3(x_1,x_2) \end{array}\right)=i
%\left( \begin{array}{cc} E_{\bf k}-E_2+i(V^2+\gamma_2)/2  & -\Omega/2 \\ -\Omega/2 & E_{\bf k}-E_2+\Delta+i\gamma_3/2  \end{array}\right)
\overleftrightarrow{\bf A}\left( \begin{array}{c} e_2(x_1,x_2) \\ e_3(x_1,x_2) \end{array}\right)-\sqrt{\f{3}{2}}i\bar{V}g(x_1,x_2,0-)\left( \begin{array}{c}  1 \\ 0 \end{array}\right), 
\label{3eq}
\eea
where $\overleftrightarrow{\bf A}$ has a similar form of the square matrix which appears in Eq.(\ref{smat}). Following the similar procedure that we have used for the two-photon case, we can solve the above equations to find the amplitudes  for all full values of  $x_1$, $x_2$ and $x_3$. Below we provide explicit expression of different amplitudes of the three-photon state.
\bea
g(x_1,x_2,x_3)&=&\f{1}{\sqrt{3!}} \Big[\sum_P g_{k_{P_1}}(x_1)g_{k_{P_2}}(x_2)g_{k_{P_3}}(x_3)+\sum_{PQ} g_{k_{P_3}}(x_{Q_3})B^{(2)}_{k_{P_1},k_{P_2}} (x_{Q_1},x_{Q_2})\theta(x_{Q_2})\nn \\
&&~~~~~~~~~~~~~~~~~~~~~~~~~
~~~~~~~~~~~~~~~~~~~~~~~~~~~~~~~+\sum_{PQ}B^{(3)}_{k_{P_1},k_{P_2},k_{P_3}} (x_{Q_1},x_{Q_2},x_{Q_3})\theta(x_{Q_3 }) \Big], 
\label{3S1}\\
e_2(x_1,x_2)&=&\f{\beta}{2\sqrt{2\pi}} \sum_P g_{k_{P_1}}(x_1)g_{k_{P_2}}(x_2)\Gamma_{k_{P_3}}(0)+
\sqrt{\f{2}{3}}\,\f{\beta}{2\sqrt{2 \pi}}\sum_{P,R} \Gamma_{k_{P_3}}(0) B^{(2)}_{k_{P_1},k_{P_2}}(x_{R_1},x_{R_2})\theta(x_{R_2}) \nn \\
&&+\f{\beta}{2} \sum_{P,R} g_{k_{P_2}}(x_{R_2}) (1-t_{k_{P_1}})
\Gamma_{k_{P_3}}(x_{R_1}) h_{k_{P_1}}(x_{R_1})h_{k_{P_3}}(x_{R_1}) \theta(x_{R_1}) \nn \\
&&+\beta \sum_{P,R}  (1-t_{k_{P_1}})(1-t_{k_{P_2}})
\Gamma_{k_{P_3}}(x_{R_1}) h_{k_{P_1}}(x_{R_1}) h_{k_{P_2}}(x_{R_1})h_{k_{P_3}}(x_{R_2}) \theta(x_{R_{12}})\theta(x_{R_2}) ,
\label{3S2}\\
 e_3(x_1,x_2)&=&\f{\beta}{2\sqrt{2\pi}} \sum_P g_{k_{P_1}}(x_1)g_{k_{P_2}}(x_2)\Upsilon_{k_{P_3}}(0)+
\sqrt{\f{2}{3}}\,\f{\beta}{2\sqrt{2 \pi}}\sum_{P,R} \Upsilon_{k_{P_3}}(0) B^{(2)}_{k_{P_1},k_{P_2}}(x_{R_1},x_{R_2})\theta(x_{R_2}) \nn \\
&&+\f{\beta}{2} \sum_{P,R} g_{k_{P_2}}(x_{R_2}) (1-t_{k_{P_1}})
\Upsilon_{k_{P_3}}(x_{R_1}) h_{k_{P_1}}(x_{R_1})h_{k_{P_3}}(x_{R_1}) \theta(x_{R_1}) \nn \\
&&+\beta \sum_{P,R} (1-t_{k_{P_1}})(1-t_{k_{P_2}})
\Upsilon_{k_{P_3}}(x_{R_1}) h_{k_{P_1}}(x_{R_1}) h_{k_{P_2}}(x_{R_1})h_{k_{P_3}}(x_{R_2}) \theta(x_{R_{12}})\theta(x_{R_2}), 
\label{3S3} \\
j(x_1,x_2;x_3)&=&\f{1}{2!} \Big[\sum_R g_{k_{R_1}}(x_1)g_{k_{R_2}}(x_2)h_{k_{3}}(x_3)+\sum_{R S}B^{(2)}_{k_{R_1},k_{R_2}} (x_{S_1},x_{S_2})\theta(x_{S_2})h_{k_3}(x_3) \Big], \\
v_2(x_1;x_3)&=& \f{\beta}{\sqrt{2 \pi}}\sum_R g_{k_{R_1}}(x_1)\Gamma_{k_{R_2}}(0) h_{k_3}(x_3) +\beta \sum_R (1-t_{k_{R_1}})\Gamma_{k_{R_2}}(x_1) h_{k_{R_1}}(x_1)h_{k_{R_2}}(x_1) \theta(x_1)h_{k_3}(x_3), \\
v_3(x_1;x_3)&=&\f{\beta}{\sqrt{2 \pi}}\sum_R g_{k_{R_1}}(x_1){\Upsilon_{k_{R_2}}(0) } h_{k_3}(x_3) +\beta \sum_R (1-t_{k_{R_1}})\Upsilon_{k_{R_2}}(x_1) h_{k_{R_1}}(x_1)h_{k_{R_2}}(x_1) \theta(x_1)h_{k_3}(x_3), \\
f(x_3;x_1,x_2)&=&\f{1}{2!} \sum_R g_{k_{3}}(x_3)h_{k_{R_1}}(x_1)h_{k_{R_2}}(x_2), ~~w_2(x_1,x_2)= \f{1}{2!} \f{\beta}{\sqrt{ 2 \pi}}\sum_R \Gamma_{k_{3}}(0)h_{k_{R_1}}(x_1)h_{k_{R_2}}(x_2), \\
w_3(x_1,x_2)&=&\f{1}{2!}  \f{\beta}{\sqrt{ 2 \pi}}\sum_R \Upsilon_{k_{3}}(0)h_{k_{R_1}}(x_1)h_{k_{R_2}}(x_2),~~h(x_1,x_2,x_3)= \f{1}{\sqrt{3!}} \sum_P h_{k_{P_1}}(x_1)h_{k_{P_2}}(x_2)h_{k_{P_3}}(x_3),\\
B^{(3)}_{k_{P_1},k_{P_2},k_{P_3}}&&(x_{Q_1},x_{Q_2},x_{Q_3})=-2i \bar{V} \beta (1-t_{k_{P_1}})(1-t_{k_{P_2}})\Gamma_{k_{P_3}}(x_{Q_{13}}) h_{k_{P_1}}(x_{Q_1})h_{k_{P_2}}(x_{Q_1})h_{k_{P_3}}(x_{Q_2})\theta(x_{Q_{12}})\theta(x_{Q_{23}}).\nn
\eea
where $P=(P_1,P_2,P_3)$, $Q=(Q_1,Q_2,Q_3)$ are permutation of $(1,2,3)$, and $R=(R_1,R_2)$, $S=(S_1,S_2)$ are the permutation of $(1,2)$. The term $B^{(3)}_{k_{P_1},k_{P_2},k_{P_3}}(x_{Q_1},x_{Q_2},x_{Q_3})$ in Eq.(\ref{3S1}) is a three-photon bound state.  The above method of constructing scattering eigenstates can be generalized to four and more incident photons. However it becomes difficult to extract measurable physical quantities from these multi-photon scattering states because of their complex structure.
\end{widetext}
\section{Second-order coherence}
\label{app2}
As we discuss in the main text we only keep higher-order contributions in the numerator and denominator of the second-order coherence. The main contribution in the numerator of the second-order coherence comes from the two-photon sector of the scattering state for a weak incident probe beam. The zero- or single-photon scattering state does not contribute in the numerator as two annihilation operators in the numerator destroy the zero- or single-photon state. There would be some contributions in the numerator from three- or more photons which we neglect for a weak incident probe beam. %However, the contribution in the numerator from the three-photon scattering state is much smaller than the contribution coming from the two-photon state. Thus we neglect it here. 
 Therefore we find by using the two-photon scattering state $| k_1,k_2 \ra$ in Eq.(\ref{wavefn}) for $|\psi\ra$ in the numerator of the second-order coherence for the scattered photons at $x_1,x_2>0$,
\begin{widetext}
\bea
\la k_1,k_2|a^{\dg}_{m}(x_1)a^{\dg}_m(x_2)a_m(x_2)a_m(x_1)| k_1,k_2 \ra=|(\frac{1}{\sqrt{2}}A_2 g(x_1,x_2)\pm\frac{1}{2}B_2(j(x_1;x_2)+j(x_2;x_1))+\frac{1}{\sqrt{2}}C_2  h(x_1,x_2))|^2,\nn\\
\eea
\end{widetext}
where $+$ sign is for the transmitted probe photons when $m=R$ and $-$ sign is for the reflected probe photons when $m=L$ for an incident state coming from the left. In the denominator of the second-order coherence, the contribution from the single-photon scattering state is much larger than that from the two or more photons. Therefore we keep only single photon contribution in the denominator. However we keep contributions from one incident photon with momentum $k_1$ and the other with momentum $k_2$. We find for the denominator for $x_1,x_2>0$, $\sum_{i,j=1,2}\la k_i|a^{\dg}_{m}(x_1)a_m(x_1)|k_i \ra \la k_j|a^{\dg}_{m}(x_2)a_m(x_2)|k_j \ra =\sum_{i,j=1,2}|(A_1 g_{k_i}(x_1)\pm B_1 h_{k_i}(x_1))(A_1 g_{k_j}(x_2)\pm B_1 h_{k_j}(x_2))|^2/4=\sum_{i,j=1,2}|(t_{k_i} \pm 1 )(t_{k_j} \pm 1)/(8\pi)|^2 \approx |(t_{k_1} \pm 1 )(t_{k_2} \pm 1)/(4 \pi)|^2$, where the $+$ and $-$ signs are respectively for $m=R$ and $m=L$. We assume $k_1 \sim k_2$ for photons of the probe beam for the last step of derivation.

\end{document}